\documentclass{article}
\usepackage{amsmath}
\usepackage{amssymb}
\usepackage{graphicx}
\usepackage{epstopdf}
\begin{document}
\title{Five dimensional cosmological traversable wormhole}
\author{S. Najafi, T. Rostami\footnote{Electronic address: t\_rostami@sbu.ac.ir} and S. Jalalzadeh\footnote{Electronic address: s-jalalzadeh@sbu.ac.ir} \\
\small Department of Physics, Shahid Beheshti University, Evin,
Tehran 19839 Iran}
\maketitle
\begin{abstract}
In this paper, a traversable wormhole in the Friedmann-Lema\^{\i}tre-Robertson-Walker (FLRW) model with one extra spacelike compact dimension is studied. We have chosen dynamical compactification as the evolution of the fifth dimension. In this respect, we study how the existence of the extra dimension, affect the behavior of the energy density, the shape function and the scale factor. It is shown that the total matter can be non-exotic and the violation of the weak energy condition can be avoided.
\end{abstract}
\maketitle
\section{Introduction}\label{sec1}
One of the controversial problems with static traversable wormholes introduced by Morris and Thorne (MT) \cite{1} is the unphysical stress-energy tensor with negative energy density. It is shown that the matter supporting this kind of structure  violates the physical laws, under which, normal matter behaves. Thus, it would be a kind of exotic matter with strange behavior that violates the energy conditions \cite{2,3}. Since then, so many generalizations have been done to build the MT wormhole with normal matter, which, does not violate the energy conditions \cite{4,5,6,7,8}. The energy conditions of general relativity are determined, in order to study the notion of ``locally" positive energy densities precisely. The energy conditions indicate that various linear combinations of the stress-energy tensor should be positive or at least non-negative \cite{visser}. The weaker assertion of the energy conditions is the averaged energy conditions, which permit localized violation of the energy conditions, in which the energy conditions holds when integrating along null or timelike geodesics \cite{tipler}.
Many attempts have been done in order to avoid violation of the energy conditions. In \cite{11}, entering wormhole in an inflating universe was suggested, and it is proven that, the more inflation goes on, the less exotic matter needed to construct the wormhole structure. Embedding of the MT wormhole in the FLRW model of the universe is investigated in \cite{12}. It is shown that, for normal matters like dust and radiation (with equation of state $\omega=0$  and $\omega=1/3$ respectively) for which, the scale factor is a power function of time, the violation of the null energy condition (NEC) can be avoided. A stationary axially symmetric rotating wormhole is described in \cite{13,14}, in which, the violation of the energy conditions will be avoided for some sets of geodesics entering wormhole. N-dimensional static and evolving Lorentzian wormhole solutions with a cosmological constant are obtained in \cite{15}.

Toward studying the existence of the wormholes, they have been studied in different alternative theories of general relativity \cite{16,17,18,19,20,21,F. Darabi}.

Furthermore, the wormhole was discussed from different points of views. For example, the conversion of a wormhole into a black hole or vise versa is discussed by Hayward in new standpoint \cite{24}, and also Refs. \cite{25,26}. Thermodynamics of wormholes treated in \cite{27}. Gravitational microlensing as an observable effect of wormholes was first proposed by Cramer et.al in \cite{29}. Since then, many authors investigated the observable effects of this structure: \cite{30,31,32,33,34}.

Along with, a great deal of interest has been focused on wormholes in higher dimensions, which arises from the theories
of fundamental physics, such as string theory, supergravity, Kaluza-Klein (KK)\cite{higher,KIM,Riazi}

In this paper, we consider a traversable wormhole in a $5$-dimensional FLRW spacetime, in which the fifth dimension is spacelike and is considered to be compact, on the basis of KK theory. As we know, KK theory is one of the first efforts to unify all the forces under one fundamental law. It was shown that by the addition of one extra spatial compact dimension which is considered to be as small as Planck's length, gravity and electromagnetism can be unified \cite{35,36,37}. We have used this concept to construct our higher dimensional traversable wormhole.
First, we consider dynamical compactification as the evolution of the fifth dimension \cite{mohammadi,39}. This assumption will ensure that the extra dimension remains compact and non-observable within the growth of the normal dimensions. The reason to consider such case is to discuss the following questions: `` if an extra dimension can induce a pressure which plays the role of exotic matter needed to support wormhole structure or not?'', or  ``under what conditions these assumptions can avoid the violation of the WEC?''.
Secondly, the extra dimension is assumed to be static. In both cases, we will show that the WEC can be satisfied for the total energy-momentum tensor, if the cosmic matter has: $(\omega\geq-1/3)$.
The organization of the paper is as follows: In Sec. II, the five dimensional cosmological traversable wormhole is introduced.
The extra dimension is assumed to be compact and its size evolves as an inverse power of the radius of our universe. Then we investigate the terms and conditions under which, the wormhole solution does not violate the WEC.
Finally, we consider the evolution of the fifth dimension to be constant and the WEC will be discussed. A discussion of our result is given in section III.
\section{five dimensional cosmological traversable wormhole}
Let us start from the $5$-dimensional KK theory with the action
\begin{align}\label{0010}
{\cal S}=\int d^{5}x\sqrt{-{\cal G}}{\cal R}+{\cal S}_{m}^{5},
\end{align}
where ${\cal R}$ is the Ricci scalar corresponding to the $5$-dimensional metric ${\cal G}_{ij}$ ($i,j=0,1,...,4$), and ${\cal S}_{m}^{5}$ is the $5$-dimensional matter action. We now use units such that $G=1$.
The metric element of the $5$-dimensional wormhole in an FLRW universe is given by
\begin{align}\label{010}
ds^2=-e^{2\Phi(r)}dt^2+R^2(t)\bigg\{\dfrac{dr^2}{1-kr^2-\dfrac{b(r)}{r}}+r^2d\Omega^2\bigg\}+B^2(t)dy^2\hspace{.15cm},
\end{align}
where $k$ denotes the spatial curvature constant with $k = -1, 0, 1$ corresponding to a closed, flat and open universe, respectively, $R$ is the scale factor for three ordinary spacelike dimensions, $B$ is the scale factor for the fifth dimension, which is considered to be spacelike, $\Phi(r)$ and $b(r)$ are arbitrary functions of the radial coordinate ($r$), denote the redshift function and the form function, respectively \cite{1}. The range of the radial coordinate increases from its minimum value at the wormhole throat $r_{0}$, to infinity. The metric is consisted of two spacetimes: FLRW universe with one extra compact dimension, and the static MT traversable wormhole metric.

A main property of a wormhole is the flare out condition at the throat, which is given by $\frac{b-b'r}{b^{2}}>0$, is imposed which will preserve the shape of a traversable wormhole \cite{1}. Furthermore, to have a wormhole solution, at the throat, $b(r_{0})=r_{0}$ the condition $b'(r_{0})<1$ is imposed.
In addition, to have a traversable wormhole there must be no event horizon which implies $\Phi(r)$ must be finite everywhere.

The Einstein's field equations for the perfect fluid with the energy-momentum tensor given by $T^{i}_{j}=diag[-\rho(t,r),-\tau(t,r),p(t,r),p(t,r),p_{y}(t,r)]$, in the case of zero radial tidal force wormhole ($\Phi(r)=0$), are given by
\begin{eqnarray}\label{013}
8\pi\rho(t,r)=-T^{t}_{t}=\dfrac{3\dot{R}^2}{R^2}+\dfrac{3\dot{R}\dot{B}}{RB}+\dfrac{1}{R^2}\bigg(3k+\dfrac{b'}{r^2}\bigg),
\end{eqnarray}
\begin{eqnarray}\label{014}
8\pi \tau(t,r)=T^{r}_{r}=\dfrac{2\ddot{R}}{R}+\dfrac{\dot{R}^2}{R^2}+\dfrac{2\dot{R}\dot{B}}{RB}+\dfrac{\ddot{B}}{B}+\dfrac{1}{R^2}\bigg(k+\dfrac{b}{r^3}\bigg),
\end{eqnarray}
\begin{eqnarray}\label{015}
8\pi p(t,r)=T^{\theta}_{\theta}=-\bigg\{ \dfrac{2\ddot{R}}{R}+\dfrac{\dot{R}^2}{R^2}+\dfrac{2\dot{R}\dot{B}}{RB}+\dfrac{\ddot{B}}{B}+\dfrac{1}{R^2}\bigg(k-\dfrac{b}{2r^3}+\dfrac{b'}{2r^2}\bigg) \bigg\},
\end{eqnarray}
\begin{eqnarray}\label{016}
8\pi p_{y}(t,r)=T^{y}_{y}=-\bigg\{ \dfrac{3\ddot{R}}{R}+\dfrac{3\dot{R}^2}{R^2}+\dfrac{1}{R^2}\bigg(3k+\dfrac{b'}{r^2}\bigg) \bigg\}.
\end{eqnarray}
Since, $\rho$, $\tau$, $p$ and $p_{y}$ depend on both $r$ and $t$, to solve the Einstein field equations, we can choose the following Ansatz
\begin{eqnarray}\label{018}
\begin{array}{cc}
R^2(t)\rho(t,r)=R^2(t)\rho_{c}(t)+\rho_{w}(r),\\ R^2(t)\tau(t,r)=R^2(t)\tau_{c}(t)+\tau_{w}(r), \\ R^2(t)p(t,r)=R^2(t)p_{c}(t)+p_{w}(r),  \\
R^2(t)p_{y}(t,r)=R^2(t)p_{y_{c}}(t)+p_{y_{w}}(r).
\end{array}
\end{eqnarray}
The subscript $c$ refers to the cosmological part and $w$ refers to the wormhole part. Indeed, this is due to the fact that we simply assume the wormhole is small compared to the universe.
With the above Ansatz, the Einstein field equations and the conservation laws are separated as
\begin{eqnarray}\label{019}
R^2 \rho_{c}(t)-\dfrac{3}{8\pi}\bigg[\dot{R}^2+\dfrac{R\dot{B}\dot{R}}{B}+k\bigg]&=-\rho_{w}(r)+\dfrac{b'}{8 \pi r^2}=l,
\end{eqnarray}
\begin{eqnarray}\label{020}
R^2 \tau_{c}(t)-\dfrac{1}{8\pi}\bigg[2R\ddot{R}+\dot{R}^2+\dfrac{2\dot{R}\dot{B}R}{B}+\dfrac{\ddot{B}R^2}{B}+k\bigg]=-\tau_{w}(r)+\dfrac{b}{8\pi r^3}=m,
\end{eqnarray}
\begin{eqnarray}\label{021}
R^2 p_{c}(t)+\dfrac{1}{8\pi}\bigg[2R\ddot{R}+\dot{R}^2+\dfrac{2\dot{R}\dot{B}R}{B}+\dfrac{\ddot{B}R^2}{B}+k\bigg]=-p_{w}(r)-\dfrac{1}{8\pi}\bigg[\dfrac{-b}{2r^3}+\dfrac{b'}{2r^2}\bigg]=\alpha,
\end{eqnarray}
\begin{eqnarray}\label{022}
R^2 p_{y_{c}}(t)+\dfrac{3}{8\pi}\bigg[R\ddot{R}+\dot{R}^2+k\bigg]&=-p_{y_{w}}(r)-\dfrac{1}{8 \pi}\dfrac{b'}{r^2}=\gamma,
\end{eqnarray}
\begin{eqnarray}\label{023}
\bigg\{\dfrac{d}{dt}(R^3\rho)+\dfrac{(2p-\tau)}{3}\dfrac{d}{dt}(R^3)  \bigg\}+R^3\dfrac{\dot{B}}{B}\bigg(\rho+p_{y}\bigg)=0,
\end{eqnarray}
\begin{eqnarray}\label{024}
\tau'+\dfrac{2}{r}(\tau+p)=0,
\end{eqnarray}
where $l$, $m$, $\alpha$ and $\gamma$ are the separation constants independent of both $t$ and $r$. The second term in the equation (\ref{023}), is from the extra dimension which modifies the $4$-dimensional energy-momentum conservation. As introduced in \cite{mohammadi}, we assume for the scale factor of the extra dimension to evolve as
\begin{eqnarray}\label{00020}
B(t)=\frac{A}{R^{n}},
\end{eqnarray}
where $A$ is a constant and the parameter $n$ must be positive in order to have dynamical compactification and non-observable extra dimension during the evolution of our universe. Thus, as the scale factor of our universe increase in size, $B$ shrinks. In addition, we consider a particular solution for which the pressure along the fifth dimension vanishes. As the simplest case, one may assume $B(t)$ to be constant, this makes the second term in (\ref{023}) vanish \cite{mohammadi}.

Now we will focus our attention to the above cases, and investigate the solutions and discuss about the conditions needed to have the non-exotic matter maintaining the wormhole in flat universe $(k=0)$.
To check the WEC, we define the exotic function as proposed in \cite{12},
\begin{eqnarray}\label{0020}
\zeta=\frac{\tau-\rho}{|\rho|}.
\end{eqnarray}
Indeed, this term plays an important role in studying the WEC condition. For the static $4$-dimensional wormhole case $\zeta$ function, is proportional to $\frac{b-b'r}{b^2}$, which must be positive or $\tau>\rho$, consequently, is exotic matter.
In what follows, we investigate the WEC in our $5$-dimensional wormhole.

\subsection{Dynamically compact extra dimension}
In this subsection we consider dynamical compactification for the evolution of the fifth dimension introduced in (\ref{00020}). Inserting this in equation (\ref{023}), allows us to define an effective pressure $\widetilde{P}$ as
\begin{eqnarray}\label{032}
\widetilde{P}\equiv \dfrac{1}{3}(2p-\tau-n\rho-np_{y}).
\end{eqnarray}
The Einstein's field equations and the conservation laws for this case are as follows
\begin{eqnarray}\label{034}
\widetilde{P}= \dfrac{1}{8\pi}\bigg[2\dfrac{\ddot{R}}{R}(n-1)&+\dfrac{\dot{R}^2}{R^2}(n-1)-\dfrac{1}{R^2}(\dfrac{b'}{3r^2}) \bigg],
\end{eqnarray}
\begin{eqnarray}\label{035}
R^2\rho_{c}(t)-\dfrac{3}{8 \pi}\bigg[\dot{R}^2(1-n)\bigg]&=-\rho_{w}+\dfrac{b'}{8 \pi r^2}=l,
\end{eqnarray}
\begin{eqnarray}\label{036}
R^2\tau_{c}(t)-&\dfrac{1}{8 \pi}\bigg[\ddot{R}R(2-n)+\dot{R}^2(n^2-n+1)\bigg]=-\tau_{w}+\dfrac{b}{8 \pi r^3}=m,
\end{eqnarray}
\begin{eqnarray}\label{037}
R^2p_{c}(t)&+\dfrac{1}{8 \pi}\bigg[\ddot{R}R(2-n)+\dot{R}^2(n^2-n+1)\bigg]=-p_{w}-\dfrac{1}{8\pi}\bigg(\dfrac{-b}{2r^3}+\dfrac{b'}{2r^2}\bigg)=\alpha,
\end{eqnarray}
\begin{eqnarray}\label{038}
R^2p_{y_{c}}(t)&+\dfrac{3}{8\pi}\bigg[\ddot{R}R+\dot{R}^2\bigg]= -p_{y_{w}}-\dfrac{b'}{8\pi r^2}=\gamma,
\end{eqnarray}
\begin{eqnarray}\label{039}
\begin{array}{cc}
\dfrac{R^3}{\dot{R}}\bigg\{ \dot{\rho_{c}}+\dfrac{\dot{R}}{R}\bigg[(3-n)\rho_{c}+3p_{c}-np_{y_{c}}\bigg] \bigg\}=\\ (n-1)\rho_{w}-2p_{w}+\tau_{w}+np_{y_{w}}=(1-n)l+2\alpha-m-n\gamma=q,
\end{array}
\end{eqnarray}
\begin{eqnarray}\label{040}
\tau_{w}'+\dfrac{2}{r}(\tau_{w}+p_{w})=0.
\end{eqnarray}
For the cosmological principle to be held, we must take the cosmic parts of our equations isotropic so $\tau_{c}(t)=-p_{c}(t)$, which gives $m=-\alpha$.
Regarding to the fact that the cosmic part is just a function of $t$ and the wormhole part is a function of $r$,
we take the equations of state for the cosmic part and the wormhole part of the energy-momentum tensor as $p_{c}(t) =\omega \rho_{c}(t) $ and $p_{w}(r) =\beta \rho_{w}(r)$ respectively. Then, using equations (\ref{035}) and (\ref{037}) in to the equation (\ref{038}), with the above equation of state for the cosmic part, one can easily obtain the cosmic pressure along the extra dimension as
\begin{eqnarray}\label{041}
p_{y_{c}}(t)&=\dfrac{-\rho_{c}}{(n-2)(1-n)}\bigg[3\omega(1-n)+n^2-1\bigg]-\bigg[\dfrac{l(n+1)-3\alpha-\gamma(n-2)}{(n-2)R^2}\bigg].
\end{eqnarray}
Substituting $p_{y_{c}}(t)$ from the above equation into the conservation law equation (\ref{039}), the cosmic energy-density becomes
\begin{eqnarray}\label{042}
\rho_{c}=\rho_{0}R^{[-6\omega(n-1)+2n^2-4n+6]/(n-2)}+\dfrac{1}{R^2}\bigg[ \dfrac{2l(n^2-n+1)-6\alpha(n-1)}{-6\omega(n-1)+2(n^2-n+1)} \bigg].
\end{eqnarray}
where $\rho_{0}$ is the integration constant. Using the similar procedure for the wormhole part, we obtain
\begin{eqnarray}\label{043}
\begin{array}{cc}
\rho_{w}=C_{1}r^{-2(\frac{1+3\beta}{1+2\beta})}-\bigg(\dfrac{l+3\alpha}{1+3\beta}\bigg),\\
\tau_{w}=\rho_{w}(1+2\beta)+l+3\alpha,\\
p_{y_{w}}(r)=-\rho_{w}-l-\gamma.
\end{array}
\end{eqnarray}
Taking into account $\rho_{w}$ obtained from the above equations, together with the equations of state for the wormhole part introduced above, into equations (\ref{035}) and (\ref{037}), we obtain for the shape function,
\begin{eqnarray}\label{1}
b(r)=8\pi C_{1}(1+2\beta)r^{(\frac{1}{1+2\beta})}+\dfrac{8\pi r^3}{1+3\beta}(l\beta-\alpha),
\end{eqnarray}
where $C_{1}$ is the integration constant and other integration constants are set to be zero by the asymptotic flatness. The asymptotic flatness condition, ($lim_{r\rightarrow\infty}\frac{b(r)}{r}=0$), implies that $\beta<\frac{-1}{2}$, $C_{1}<0$, $\alpha=0$ and $l=0$. In addition, the argumentations mentioned below equation (\ref{040}), directly results in $m=0$.
The flare-out condition is automatically satisfied with the above restrictions on the constants
\begin{eqnarray}\label{044}
\bigg(\dfrac{b-b'r}{b^2}\bigg)=\dfrac{\beta}{4\pi C_{1} (1+2\beta)^2}r^{-\frac{1}{1+2\beta}}>0.
\end{eqnarray}
The above condition is satisfied for the entire range of $r_{0}<r<\infty$.
On the other hand, at the throat we have $b(r_{0})=r_{0}$, this leads us to
\begin{eqnarray}\label{045}
8\pi C_{1}(1+2\beta)=r_{0}^\frac{2\beta}{1+2\beta}.
\end{eqnarray}
Hence, the constants $\beta$ and $C_{1}$ are related to the throat radius $r_{0}$.
Using (\ref{042}) in the equation (\ref{035}), one can obtain for the scale factor
\begin{eqnarray}\label{046}
R=R_{0}\bigg(\dfrac{t}{t_{0}} \bigg)^\frac{n-2}{3\omega(n-1)-n^2+2n-3}.
\end{eqnarray}
where $R_{0}$ is the integration constant and $t_{0}=\frac{3\omega(n-1)-n^2+2n-3}{n-2}\sqrt{\frac{8\pi}{3}\rho_{0}}$. Inserting the equation (\ref{046}) into (\ref{042}), the cosmic energy-density behaves as
\begin{eqnarray}\label{047}
\rho_{c}\sim t^{-2}.
\end{eqnarray}
The WEC is satisfied when the following inequalities hold simultaneously
\begin{eqnarray}\label{048}
\begin{array}{cc}
\rho=\dfrac{1}{8\pi}\bigg\{\dfrac{3(1-n)\dot{R}^2}{R^2}+\dfrac{1}{R^2}\dfrac{b'}{r^2}  \bigg\}\geq0,\\
\zeta=\dfrac{1}{8\pi \mid\rho\mid}\bigg\{\dfrac{\ddot{R}}{R}(2-n)+\dfrac{\dot{R}^2}{R^2}(n^2+2n-2) +\dfrac{b^2}{r^3R^2}\bigg(\dfrac{b-b'r}{b^2}\bigg)
 \bigg\}\leq0,\\
\rho+p_{y}=\dfrac{-3}{8\pi} \bigg(\dfrac{\ddot{R}}{R}+n\dfrac{\dot{R}^2}{R^2}  \bigg)\geq0.
\end{array}
\end{eqnarray}
Substituting equations (\ref{046}) and (\ref{1}) in the above inequalities, lead to
\begin{eqnarray}\label{049}
\begin{array}{cc}
\rho=\dfrac{3(1-n)t^{-2}}{8\pi}\left[\frac{n-2}{3\omega(n-1)-n^2+2n-3}\right]^2+\frac{8\pi C_{1}}{R^2}r^{-2\left(\frac{1+3\beta}{1+2\beta}\right)}\geq0,\\
\zeta=\dfrac{1}{8\pi\mid\rho\mid}\bigg\{3(n-1)(\omega+1)t^{-2} \left[\dfrac{n-2}{3\omega(n-1)-n^2+2n-3}\right]^2+\dfrac{16\pi C_{1}\beta}{R^2}r^{-2\left(\frac{1+3\beta}{1+2\beta}\right)} \bigg\}
\leq0,\\
\rho+p_{y}=\dfrac{(n-2) t^{-2}\bigg\{ 9\omega(n-1)-6n^2+9n-3 \bigg  \}}{8\pi[3\omega(n-1)-n^2+2n-3]^2}\geq0.
\end{array}
\end{eqnarray}
The first inequality holds with $n<1$, when $C_{1}$ at the throat have the value of
\begin{eqnarray}\label{050}
C_{1}=\frac{n-1}{64\pi^2}(\frac{R_{0}}{t_{0}})^2 \left[\dfrac{n-2}{3\omega(n-1)-n^2+2n-3}\right]^2r_{0}^{2\left(\frac{1+3\beta}{1+2\beta}\right)},
\end{eqnarray}
This means, that the WEC will be satisfied within the interval $r_{0}\leq r<\infty$. Now, by inserting equations (\ref{050}) and (\ref{045}) in the second inequality of the equation (\ref{049}), it is also satisfied within the above mentioned interval for the radial coordinate, if only we have $\omega>-1$, in constant time $t_{0}$.

\begin{figure}
  \centering
   \hspace{1.cm}\includegraphics[width=8cm]{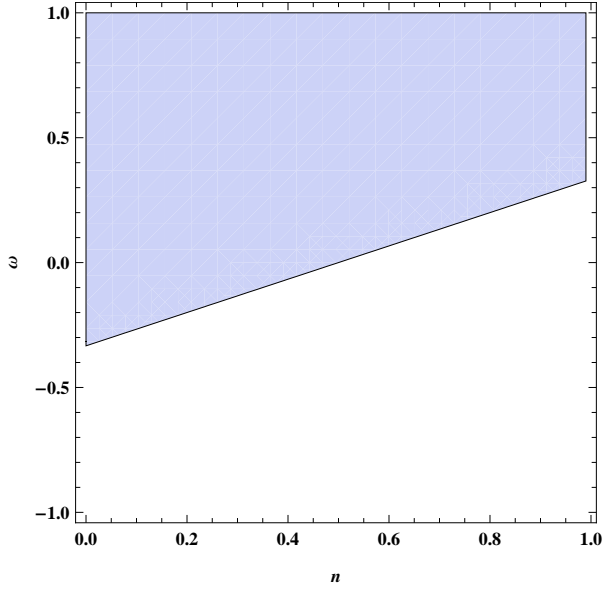}\\
  \caption{The behavior of $\rho+p_{y}\geq 0$ for $t_{0}=1$}\label{fig1}
\end{figure}

In order to see whether the WEC is satisfied we have to check the third inequality of the equation (\ref{049}). It is held if
\begin{eqnarray}\label{0050}
\begin{cases}
-\dfrac{1}{3}\leq\omega<0, &(0\leq n\leq \dfrac{3\omega+1}{2}),\\
 0\leq\omega<\dfrac{1}{3}, &(0\leq n\leq \dfrac{1}{2}),\\
\omega\geq\dfrac{1}{3}, &(0\leq n<1).
\end{cases}
\end{eqnarray}
Figure \ref{fig1} better illustrates the third inequality of the equation (\ref{049}), plotted with constant time $t_{0}=1$. The figure shows that, for $0\leq n<1$, $\omega$ covers $\frac{-1}{3}\leq \omega<1$.
So all the above inequalities are satisfied simultaneously for $n<1$ and $\omega\geq-1/3$. Obviously, in this case, it is possible to avoid the violation of the WEC.

Now, as a particular solution, let us examine the case where the pressure of the fifth dimension is taken to be zero.
By putting the pressure along the extra dimension $p_{y}$, equal to zero in the equation (\ref{038}), the shape function can be found
\begin{eqnarray}\label{051}
b(r)=-8\pi \gamma \dfrac{r^3}{3}+c',
\end{eqnarray}
where $c'$ is the integration constant. Using equation (\ref{051}) into the equations (\ref{035}), (\ref{036}) and (\ref{037}), we obtain for the wormhole part
\begin{eqnarray}\label{052}
\begin{array}{cc}
\rho_{w}=-(l+\gamma),\\ \tau_{w}=-m-\dfrac{\gamma}{3}+\dfrac{c'}{8\pi r^3},\\ p_{w}=-\alpha-\dfrac{2\gamma}{3}+\dfrac{c'}{16\pi r^3}.
\end{array}
\end{eqnarray}
From the equation (\ref{051}), asymptotically flatness condition implies that $\gamma=0$. Therewith, the shape function at the throat reduces to $b(r_{0})=r_{0}=c'$.
From the equation (\ref{038}), the behavior of the scale factor can be obtained
\begin{eqnarray}\label{055}
R=\pm\sqrt{2D_{1}t+2D_{2}},
\end{eqnarray}
in which, $D_{1}$ and $D_{2}$ are the integration constants. In this case, the WEC is satisfied if the beneath inequalities hold simultaneously
\begin{eqnarray}\label{056}
\begin{array}{cc}
\rho=\dfrac{3}{8\pi}\bigg[ \dfrac{D_{1}^2 (1-n)}{4[D_{1}t+D_{2}]^2}  \bigg]\geq0,\\ \zeta=\dfrac{1}{8\pi\mid\rho\mid}\bigg\{\dfrac{-3(1+\omega)D_{1}^2(1-n)}{4[D_{1}t+D_{2}]^2}+\dfrac{b^2}{r^3R^2}\bigg(\dfrac{b-b'r}{b^2}\bigg)\bigg\}\leq0.
\end{array}
\end{eqnarray}
The first inequality, holds everywhere if $n<1$. The second term in $\zeta$ function is the flare out condition, which is positive and is equal to $\frac{c'}{R^2r^3}$. However, for the second inequality to be held, $r$ should be bigger than $\frac{R_{0}}{\left[3D_{1}^2(1+\omega)(1-n)\right]^{1/2}}=r_{0}$, in constant time $t_{0}$. Accordingly, it is possible to avoid the violation of the WEC provided that $\omega>-1$ with $n<1$, which supports both normal matter and dark energy. But, since the scale factor has the decelerating behavior, as is obvious from the equation (\ref{055}), this is clearly in contrast to the existence of dark energy. Consequently, the WEC condition is satisfied for the $\omega>\frac{-1}{3}$ with $n<1$.
\subsection{Static extra dimension}
In this subsection, the scale factor of the extra dimension is taken to be constant with time.
The Einstein's field equations and the conservation laws for this case are as follows
\begin{eqnarray}\label{00035}
R^2\rho_{c}(t)-\dfrac{3}{8 \pi}\dot{R}^2=-\rho_{w}+\dfrac{b'}{8 \pi r^2}=l,
\end{eqnarray}
\begin{eqnarray}\label{00036}
R^2\tau_{c}(t)-\dfrac{1}{8 \pi}\bigg[2\ddot{R}R+\dot{R}^2\bigg]=-\tau_{w}+\dfrac{b}{8 \pi r^3}=m,
\end{eqnarray}
\begin{eqnarray}\label{00037}
R^2p_{c}(t)&+\dfrac{1}{8 \pi}\bigg[2\ddot{R}R+\dot{R}^2\bigg]=-p_{w}-\dfrac{1}{8\pi}\bigg(\dfrac{-b}{2r^3}+\dfrac{b'}{2r^2}\bigg)=\alpha,
\end{eqnarray}
\begin{eqnarray}\label{00038}
R^2p_{y_{c}}(t)+\dfrac{3}{8\pi}\bigg[\ddot{R}R+\dot{R}^2\bigg]= -p_{y_{w}}-\dfrac{b'}{8\pi r^2}=\gamma,
\end{eqnarray}
\begin{eqnarray}\label{00039}
\begin{array}{cc}
\dfrac{R^3}{\dot{R}}\bigg\{\dot{\rho_{c}}+\dfrac{\dot{R}}{R}\bigg[3\rho_{c}+3p_{c}\bigg]\bigg\}=\\ -\rho_{w}-2p_{w}+\tau_{w}=l+2\alpha-m=q,
\end{array}
\end{eqnarray}
\begin{eqnarray}\label{00040}
\tau_{w}'+\dfrac{2}{r}(\tau_{w}+p_{w})=0.
\end{eqnarray}
Assuming the cosmological principle is held, we must take the cosmic parts of our equations isotropic so $\tau_{c}(t)=-p_{c}(t)$, which gives $m=-\alpha$.
With the same procedure as in the previous subsection, we take the equations of state for the cosmic part and the wormhole part of the energy-momentum tensor as $p_{c}(t) =\omega \rho_{c}(t) $ and $p_{w}(r) =\beta \rho_{w}(r)$ respectively. The components of the energy-momentum tensor for the cosmic part and the wormhole part, the scale factor and the shape function can be found
\begin{eqnarray}\label{058}
\rho_{c}=\rho_{0}R^{-3(1+\omega)}+\dfrac{q}{(3\omega+1)}R^{-2},
\end{eqnarray}
\begin{eqnarray}\label{059}
p_{y_{c}}=\rho_{c}\bigg(\dfrac{3\omega-1}{2}\bigg)-\dfrac{1}{R^2}\bigg(\dfrac{3\alpha}{2}-\dfrac{l}{2}-\gamma\bigg),
\end{eqnarray}
\begin{eqnarray}\label{060}
\tau_{w}=\rho_{w}(1+2\beta)+l+3\alpha,~~~~~p_{y_{w}}=-\rho_{w}-l-\gamma,
\end{eqnarray}
\begin{eqnarray}\label{062}
b(r)=8\pi C_{1}(1+2\beta)r^{(\frac{1}{1+2\beta})}+\dfrac{8\pi r^3}{1+3\beta}(l\beta-\alpha),
\end{eqnarray}
where $\rho_{0}$ and $C_{1}$, are the integration constant. The asymptotic flatness condition results in $l=0$, $\alpha=0$, $\beta<\frac{-1}{2}$ and $C_{1}<0$. Further, $m=0$, according to the discussion mentioned blow equation (\ref{00040}), also due to the equation (\ref{00039}), it leads to $q=0$.  It can be easily checked that the flare out condition is satisfied with the above mentioned restriction on the constants.
In this case, the minimum value of the $r$ coordinate, $b_{0}$ at the throat of the wormhole is obtained as
\begin{eqnarray}\label{0061}
8\pi C_{1}(1+2\beta)=r_{0}^\frac{2\beta}{1+2\beta}.
\end{eqnarray}
Thus, the constants $\beta$ and $C_{1}$, are dependent to the throat radius, through the above equation.
Accordingly, the solution of the scale factor is
\begin{eqnarray}\label{061}
R=R_{0}\bigg(\dfrac{t}{t_{0}}\bigg)^{\frac{2}{3(1+\omega)}},
\end{eqnarray}
where $R_{0}$ is the integration constant and $t_{0}=\frac{3(1+\omega)}{2}\sqrt{\frac{8\pi}{3}\rho_{0}}$.

The WEC requires
\begin{eqnarray}\label{063}
\begin{array}{cc}
\rho=\rho_{0}R^{-3(1+\omega)}+\dfrac{8\pi C_{1}}{R^2}r^{-2(\frac{1+3\beta}{1+2\beta})}\geq0,
\\ \zeta=\dfrac{1}{8\pi\mid\rho\mid}\bigg\{2(\dfrac{\ddot{R}}{R}-\dfrac{\dot{R}^2}{R^2})+\dfrac{2b^2}{r^3R^2}\bigg(\dfrac{b-b'r}{2b^2}\bigg) \bigg\}\leq0, \\
\rho+p_{y}=\dfrac{-3}{8\pi}\dfrac{\ddot{R}}{R}\geq0.
\end{array}
\end{eqnarray}
The substitution of the scale factor (\ref{061}), will result in the following inequalities
\begin{eqnarray}\label{064}
\begin{array}{cc}
\rho=\rho_{0}\dfrac{t^{-2}}{6\pi(1+\omega)^2}+\dfrac{8\pi C_{1}}{R^2}r^{-2(\frac{1+3\beta}{1+2\beta})}\geq0, \\ \zeta=\dfrac{1}{8\pi\mid\rho\mid}\bigg\{\dfrac{-4t^{-2}}{3(1+\omega)}+\dfrac{16\pi C_{1} \beta}{R^2}r^{-2\frac{1+3\beta}{1+2\beta}}
 \bigg\}\leq0,\\
\rho+p_{y}=\dfrac{(1+3\omega)t^{-2}}{12\pi(1+\omega)^2}\geq0.
\end{array}
\end{eqnarray}
It is interesting to note that the first inequality holds within the interval $r_{0}\leq r<\infty$, for the $C_{1}$ to be
\begin{eqnarray}\label{0064}
C_{1}=\frac{-\rho_{0}}{8\pi}R_{0}^{-\left(1+3\omega\right)}r_{0}^{2(\frac{1+3\beta}{1+2\beta})}.
\end{eqnarray}
Using equations (\ref{0061}) and (\ref{0064}), the second inequality of (\ref{064}) is also satisfied with $\omega>-1$.
However the third inequality of (\ref{064}) is held for $\omega\geq-1/3$. Clearly, it is possible for the cosmic matter to avoid the violation of the WEC, if $\omega\geq-1/3$.


\section{Conclusion}
In this paper, we have considered a traversable wormhole in a $5$-dimensional flat FLRW spacetime. The extra dimension is considered to be spacelike and compact on the basis of the KK theory. Since in static wormholes the energy conditions are violated, the wormholes in the FLRW spacetime are being studied, which found the total matter can be non-exotic \cite{12}. In this regard, the existence of an extra dimension may help to have wormhole solution, that respect the energy conditions. The Einstein field equations of the $5$-dimensional KK theory, for the perfect fluid in the case of a zero radial tidal force wormhole are obtained. We suppose that the wormhole size is very small compared to the universe. This lets us to decompose the energy-momentum tensor of the $5$-dimensional perfect fluid, to the cosmic part and the wormhole part. Then, to solve the Einstein equation, we consider two cases of dynamically compactification for the evolution of the extra dimension introduced in \cite{mohammadi}, and the static the extra dimension. As a particular solution, we dealt with the dynamically compact extra dimension with vanishing component of the pressure along extra dimension. The WEC is investigated within the mentioned cases.
It is shown that the total matter supporting a $5$-dimensional traversable wormhole in the flat FLRW universe can be non-exotic provided that, for the equation of state to be $\omega\geq\frac{-1}{3}$, for the above cases.

\section{Acknowledgment}
The authors would like to thank the anonymous referee for their valuable comments and suggestions to improve the
quality of the paper. S. Najafi is also grateful to Dr. N. Riazi for useful conversation.

\end{document}